\documentclass[preprint,12pt]{elsarticle}

\usepackage{amssymb}
\usepackage{amsmath}
\usepackage{color}
\usepackage{longtable}
\usepackage{hyperref}
\usepackage{multirow}
\usepackage{soul}
\usepackage{bm}
\usepackage{enumitem}
\usepackage{rotating}
\usepackage{array}
\usepackage{makecell}
\usepackage{xr}

\graphicspath{{figures/},{.}}

\begin{document}

\begin{frontmatter}



\title{Predicting popularity of EV charging infrastructure from GIS data\tnoteref{label1}}

\author[1]{Milan~Straka}
\ead{milan.straka@fri.uniza.sk}

\author[2]{Pasquale~De~Falco}
\author[3]{Gabriella~Ferruzzi}
\author[3]{Daniela~Proto}
\author[4]{Gijs~van~der~Poel}
\author[1]{Shahab~Khormali}
\author[1]{\v{L}ubo\v{s} Buzna}

\cortext[cor1]{Corresponding author.}

\address[1]{University of \v{Z}ilina, Univerzitn\'{a} 8215/1, \v{Z}ilina, Slovakia}
\address[2]{Department of Engineering, University of Napoli Parthenope, Naples, Italy}
\address[3]{Department of Information Technology and Electrical Engineering, University of Napoli Federico II, Naples, Italy}
\address[4]{ElaadNL, Utrechtseweg 310 (bld. 42B), 6812 AR Arnhem (GL), The Netherlands} 

\begin{abstract}
The availability of charging infrastructure is essential for large-scale adoption of electric vehicles (EV). Charging patterns and the utilization of infrastructure have consequences not only for the energy demand, loading local power grids but influence the economic returns, parking policies and further adoption of EVs. We develop a data-driven approach that is exploiting predictors compiled from GIS data describing the urban context and urban activities near charging infrastructure to explore correlations with a comprehensive set of indicators measuring the performance of charging infrastructure. The best fit was identified for the size of the unique group of visitors (popularity) attracted by the charging infrastructure. Consecutively, charging infrastructure is ranked by popularity. The question of whether or not a given charging spot belongs to the top tier is posed as a binary classification problem and predictive performance of logistic regression regularized with an $\mathit{l}_1$ penalty, random forests and gradient boosted regression trees is evaluated. Obtained results indicate that the collected predictors contain information that can be used to predict the popularity of charging infrastructure. The significance of predictors and how they are linked with the popularity are explored as well. The proposed methodology can be used to inform charging infrastructure deployment strategies. 
\end{abstract}

\begin{graphicalabstract}
\includegraphics[width=1.0\textwidth]{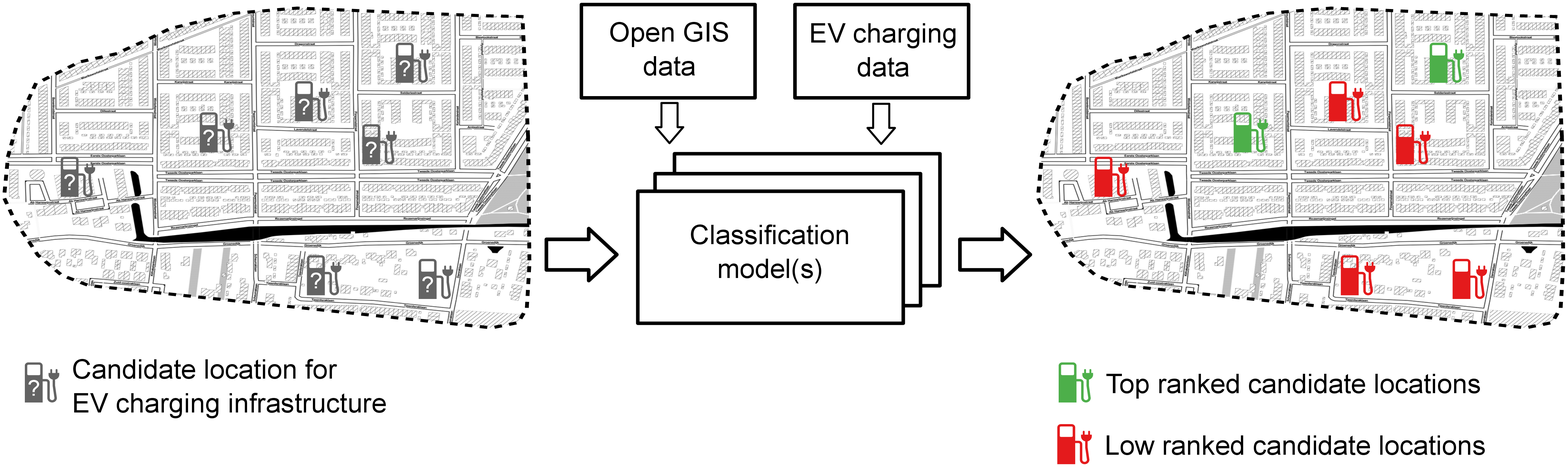}
\end{graphicalabstract}

\begin{highlights}
\item Economic and urban context predict popularity of charging infrastructure.

\item $l_1$-regularized logistic regression modestly outperforms decision trees.

\item Popularity is positively linked with charging infrastructure specs, hotels, food and financial services.

\item Popularity is negatively linked with residential areas and distance from frequently visited venues.
\end{highlights}


\end{frontmatter}

\section{Introduction}
\label{sec:intro}
Energy consumption has increased outstandingly in the last years and it will continue to be a significant global challenge. The largest portion of the total energy consumption is in the transportation sector where high economic and population growth causes a rapid increase in energy demand with excessive CO\textsubscript{2} emissions and energy crisis~\cite{EIA2013}. To mitigate the impact of the emissions of greenhouse gases and to increase energy security, automotive powertrain electrification in transportation sector could play an important role~\cite{Tali_2013, wilbanks2010}. This is confirmed by the fact that several countries have announced an ambition to stop selling vehicles fueled by diesel or petrol. For example, the UK set as the target year 2040. The EU has taken a decisive step forward in implementing the EU’s commitments under the Paris Agreement for a binding domestic CO\textsubscript{2} reduction of at least 40\% until 2030~\cite{JRC_web}. Electric vehicles (EVs), as a key element of clean and green travel mode, are spreading all over the world rapidly. For instance, an ambitious target of having over 20 million EVs on the roads by 2020 has been set by the U.S. Department of Energy~\cite{EIA2013}. However, the adoption of EVs on a large scale is supposed to bring both challenges and opportunities from technical and economic points of view~\cite{ajanovic2016dissemination,Flammini2017}.

\subsection{Motivation}
\label{sec:motivation}
In order to boost EV popularity, many challenges need to be addressed.
The chicken-egg problem in the form of charging infrastructure versus EV adoption has been recognized as an important challenge restraining the growing EV ecosystem \cite{van2013data}. Currently, insufficient charging infrastructure is a significant factor that prevents larger penetration of EVs~\cite{Coffman2017}. The drivers are hesitating to buy an EV if there is not sufficient charging infrastructure, and similarly charging infrastructure operators do not invest while the number of EVs is low and not profitable. In recent years, demand-driven and strategic rollouts (i.e. strategically covering the geographic space by chargers) have been applied~\cite{Helmus2018}. The opening of a new public charging infrastructure involves the estimation of the visitation patterns to ensure as high as possible utilization to justify the allocated resources.
Prior to opening a new public charging infrastructure, the expected utilization should be estimated in order to justify the allocated resources. Hence, one way how to support decision making in this area is to develop data-driven approaches with a predictive power.

\subsection{Previous relevant work}
\label{sec:previous_relevant_work}
Different methods, such as mathematical programming, computer-based simulation and statistical learning, have been proposed to deploy EV charging infrastructure (EVCI). Mathematical programming models for the optimal location of EVCI have been proposed in several papers. The methodologies are based on the minimization of objective functions, addressing infrastructure development costs~\cite{Liu2012, Sadeghi2014}, social costs~\cite{ren2019location}, driving range and driver habits~\cite{Gonzalez2014,Dong2014,He2018}, traffic flow data~\cite{Efthymiou2017}, unmet demand~\cite{Chen2013}, and quality of service~\cite{Davidov2017}. In \cite{GE2012} was addressed the problem considering road network and distribution system network capacity constraints. The work of Asamer et al.~\cite{Asamer2016} provides a formulation of the optimal location problem for a specific category of vehicles (taxi service). A comprehensive review of different optimization techniques for EVCI can be found~\cite{Rahman2016}. 

In contrast to the large number of studies that applied mathematical programming and computer-based simulations, there are few works in the context of location analysis that are based on data analysis methods. The predictive power of various machine learning approaches (supervised regression, decision trees, support vector regression and pairwise ranking approach) to determine the ranking of potential localities for retail stores was tested in~\cite{Karamshuk2013}, concluding that geographic and mobility features strongly improve the results. The paper~\cite{Chen2015} proposed the method for prediction of bike-sharing stations utilization using data on demographics, human activity, and area function as important factors influencing optimal placement of bike-sharing stations. Two-phase feature selection method to recognize useful features (derived from heterogeneous urban open data) for bike trip demand prediction is applied. In~\cite{DSilva2018}, it was demonstrated that similarity of urban neighborhoods and localities, and spatio-temporal features can be exploited to predict successfully the temporal activity patterns of new business venues using the k-nearest neighbor method and Gaussian processes.  

In the electric vehicle domain, often, mobility data are used to estimate future demand. In~\cite{De2015}, GPS driving databases and data mining were used to plan EVCI. Real-world driving and parking events are examined to assess suitable locations for charging spots based on existing points of interest databases and a minimum-distance criterion. The efficient distribution of selected public charging points is investigated through discharge rate of EVs in~\cite{tao2018data}. The work~\cite{Yang2017} introduced a data-driven optimization-based approach for the siting and sizing of electric taxi charging infrastructure in the city to minimize the infrastructure investments.

More recently, research works based on EV charging data have started to appear in the literature. A data-driven approach to extract useful information from EV charging events was suggested in Ref.~\cite{Xydas2016}. The proposed framework combines data pre-processing, data mining and fuzzy based models with real charging events data and weather data from three counties in the UK to characterize the charging demand of electric vehicles. Four well-known data mining algorithms, namely classification and regression trees (CART), Random forest algorithm (RFA), k-nearest neighbor (k-NN), and general chi-square automatic interaction detector (CHAID), were applied in~\cite{Verma2015}. The developed approach aims to identify and classify households with EVs by analyzing their energy consumption patterns. A data-driven statistical approach to extend the EVCI is noted in~\cite{Pevec2018}. This study suggested enriching EV charging datasets with contextual information (e.g. point of interest and driving distances between charging sites) to deploy new charging infrastructures. Hence, the benefits of geographical, demographic and economic data to optimal planning of EVCI have been acknowledged by some studies. 

Realistic planning of EVCI can be achieved only if real-world data are available and accurate models are recognized. Several research studies in the field refer to the EVnetNL dataset, one of the biggest datasets available for the area of the Netherlands~\cite{elaadnl}. Contribution focusing on EV load forecasting by comparing time series (SARIMAX model) and machine learning approaches (Random Forest, Gradient Boosting Regression Trees) was presented in Ref.~\cite{Buzna2019}. Authors in~\cite{Pevec2018} built EVCI utilization prediction models combining the EVnetNL dataset with business data, such as historical data about EV charging transactions and information about competitors in the market. The EVnetNL dataset was used to investigate and compare the performance of strategic and demand-driven rollout strategies for EVCI in the Netherlands. The obtained results in this study indicate that the proper rollout strategy depends on the maturity of the market (EV-adoption) and technology (battery capacity)~\cite{Helmus2018}. The EVnetNL dataset has been studied also in~\cite{Develder2016} to analyze EV charging flexibility as demand response potential and in~\cite{Lucas2019} to test the ability of regression algorithms to predict EV charging idle time. A set of eight indicators (e.g. EV energy demand, spatial density of EV chargers, etc.) was applied to the EVnetNL dataset to assess EVCI across countries~\cite{Lucas2018}. In addition, proposed indicators are used to assess the impact of relevant public policies on the rollout and utilization of EVCI. The work~\cite{Flammini2019} analyzed the EVnetNL dataset of 400,000 EV charging transactions in the Netherlands for the year 2015 by using a weighted affine combination of beta distributions to estimate the multimodal probability distribution of charge time, connected time and idle time. By analyzing 390k transactions, the potential to shave the energy consumption peak was investigated, while using clustering techniques to categorize charging sessions by the arrival time, departure time and idle time~\cite{Sadeghianpourhamami2018}.

\subsection{Our contribution}
\label{sec:our_contribution}
From the literature, it becomes apparent that planning of the EVCI requires an interdisciplinary approach that combines energy planning and management, economic and policy considerations, social science, geography, and data science. Therefore, our work is focused on an analytical framework that captures social, demographic, urban and transport characteristics to inform strategies for EVCI deployment and to optimize the utilization of the charging infrastructure. Many previous studies have optimized the placement of EVCI by minimizing the investments while ensuring a certain level of coverage of expected demand. The drawback of solely minimizing costs may lead to a design that
is not corresponding to optimal usage patterns and hence will not generate sufficient return of investments. The popular sites might be associated with higher rollout costs but are more likely to return the investments and pay the maintenance costs or even be profitable. To our best knowledge, this is the first paper where a prediction model attempting to forecast the popularity of charging infrastructure is presented and validated. The main contributions of this paper are as follows:

\begin{itemize}
	\item We summarize goals pursued by stakeholders involved in the charging infrastructures development while selecting and describing the set of performance indicators approximating their intentions.
    \item We provide an analytical framework that captures social, demographic, urban and transport characteristics and the availability of charging opportunities of urban neighborhoods surrounding the charging infrastructure.
    \item	We evaluate the predictive power of three prediction models: logistic regression regularized with an $l_1$ penalty, random forests and gradient boosted regression trees.
    \item From results, we evaluate the significance of factors affecting the popularity of charging infrastructure.
\end{itemize}
The rest of the paper is organized as follows. More details on the EVnetNL dataset, used GIS data and selection of response variable are given in Section~\ref{sec:materials}. Section~\ref{sec:methods} presents prediction methods, training, and validation of models. Obtained results including settings of parameters, measures of predictability and characteristics affecting the popularity of public charging infrastructure are documented in Section~\ref{sec:results}. Discussion and summary of conclusions are provided in Section~\ref{sec:discussion_and_conslusions}.

\section{Materials}
\label{sec:materials}
\subsection{Charging pools}
\label{sec:charging_pools}
Recently, a document suggesting unified terminology to be used in the electromobility field was published~\cite{EVdefinitions}. It defines a charging station as a physical object with one or more charging points sharing a common user identification interface. A charging point is an energy delivery device that might have one or several connectors, where only one can be used at the same time to charge an EV. A charging pool consists of one or multiple charging stations and the associated parking lots have one operator and a single address. Charging stations located close to each other have the same underlying geographical context and hence cannot be differentiated based on GIS data. In this paper, as an object of study, the charging pools are considered.  

\subsection{EVnetNL dataset}
The EVnetNL dataset consists of more than one million charging transactions, performed on around 1~700 charging pools, distributed across the whole area of the Netherlands, by more than 50~000 EV users. Each transaction is initiated by plugging in and terminated by plugging out the EV. Each transaction is characterized by consumed energy, charging and connection time, unique identifier of a charging station and linked to EV user by RFID card. Data records span January 2012 to March 2016. The maximum available charging power at charging stations is 11~kW supporting merely slow charging. Transactions taking place in 2015 are considered in the analysis, as it is the last complete year in the dataset and the number of charging stations was already saturated. A more detailed description of the dataset can be found in Section~\ref{secSI:EVnetNL_dataset} of the Supplementary Information (SI) file.

\subsection{GIS datasets}
Open GIS data were collected from various sources to model the urban context and human activities near charging pools. Brief descriptions of used datasets is provided in Table~\ref{tab:GISdata}. Datasets are available in various formats, hence, different predictor extraction techniques, described in Section~\ref{secSI:EVnetNL_dataset} of the SI file, were required. The extracted predictors have been pre-processed using workflow derived from~\cite{kuhn2013applied,james2013introduction} and detailed in Section~\ref{secSI:data_pre_processing} of the SI file.

	\begin{longtable}[h]{p{.2\textwidth} p{.7\textwidth} p{.1\textwidth} } 
		\hline
		Dataset & Brief description & Source\\
		\hline
		Population cores & Detailed population data organized by morphologically continuous areas. & \cite{popcores}  \\
		Neighborhoods & Population data aggregated to neighborhoods. &  \cite{neighbpop}\\
		Land use & Land use data modeled by high resolution heterogeneous polygons and divided into 25 categories. & \cite{lccbs} \\ 
		Energy & Aggregated gas and electricity consumption of companies and households estimated per neighborhoods. & \cite{energyatlas} \\
		Liveability & General index describing quality of living in 5 categories (housing, residents, services, safety, and living environment) at the level of neighborhoods.  & \cite{liveability} \\
		Traffic flows & Database of traffic volumes of cars, buses and trucks on individual roads. & \cite{traffic_flows}\\
		LandScan & Ambient population density in the raster format. & \cite{landscan}\\ 
		OpenStreetMap & OpenStreetMap amenities taking form of the point data. & \cite{osm} \\
		Charging pools 2015 & Available locations of charging pools in 2015. & \cite{ocm,opp} \\
		\hline
		\caption{Overview of collected GIS datasets.}
		\label{tab:GISdata}
	\end{longtable}

\subsection{Selection of response variable: The popularity of charging pools}
\label{sec:response_variable}
To properly inform the planning and deployment of charging infrastructure, there are many interdependent and often contradicting aspects that should be considered when quantifying the performance of charging pools:

\begin{itemize}
 \item Being motivated by the need to innovate the road transport that is based on fossil fuels, from the perspective of governments, policymakers, and municipalities, one of the main goals when developing charging infrastructure is to stimulate wide use of EVs and to invest public resources efficiently and fairly.  
\item At present, the main concern of grid operators is the stability of supply systems and seamless integrations of renewable energy sources. Technologies, such as smart charging, should help to harvest the potential of EVs in reaching this goal. This requires alignment between the presence of energy produced from renewables and the charging behavior.
\item When building or extending the network, charging infrastructure operators are seeking a profit. This requires placing charging pools in locations that can generate sufficient revenues. 
\end{itemize}

A comprehensive set of performance indicators to compare EVCI rollout among continents, countries, and regions was proposed in~\cite{Lucas2018}. Having in mind different views of policymakers, municipalities, power system operators and charging infrastructure operators and considering the availability of data, we designed the following set of indicators to quantify the performance of charging pools:
\begin{itemize}
\item \textit{Consumed energy:} Dependent on the subscription program, EV drivers either pay regularly a fixed amount and have unlimited access to charging services or they are charged a fee when connecting the vehicle and pay a certain rate per unit of consumed energy. Hence, consumed energy on a charging pool is a proxy of profitability, moreover, it also indicates how difficult it is to integrate a charging pool to the power grid.

\item \textit{Number of charging transactions:} The higher the number of EVs visiting a charging pool, the higher the potential profit. On the one hand, a high number of transactions is loading the supply network more, one the other hand it can create more opportunities for smart charging.

\item \textit{Popularity of a charging pool:} The ability of a charging pool to attract a large group of EV drivers can be approximated by the number of unique RFID cards used on a pool. Popular pools can be more robust concerning random fluctuations in the usage compared to charging pools that are highly used but only by a small group of EV drivers. Moreover, public investments into popular charging pools can be considered as socially fairer. 

\item \textit{Charging time:} Complementary information about the usage of a charging pool is provided by the overall charging time (i.e. the time that is used to transfer energy between charging infrastructure and vehicle). Long charging time indicates a high utilization of a charging pool, but it can also indicate that the rated power of a charging pool should be increased.

\item \textit{Charging ratio:} An issue occurs when EV drivers tend to leave the vehicle at the charging pool for a much longer time than is required for charging. This behavior can be motivated by free parking at charging pools in cities, and it limits the access to charging spots for other EV users. Charging ratio is a number between $0$ and $1$ that is calculated by dividing the length of the time intervals that vehicles are charging with the length of the time intervals vehicles are connected to a given charging pool. On one hand, a charging pool that has a high charging ratio can reach higher profit as it is better utilized by EV drivers. On the other hand, a low charging ratio means a higher potential for smart charging.
  
\item \textit{Use-time ratio:} If the occupancy of a pool is high, other EV drivers are often forced to search for charging alternatives, which is decreasing the perceived quality of service provided by the charging pools. A simple measure, estimating the occupancy of a charging pool is the use-time ratio~\cite{Lucas2018}, which is calculated by dividing the length of the time period when the pool was occupied by the length of the observed period.

\item \textit{Energy ratio:} How much a charging pool is loading the supply network, can be estimated by the energy ratio~\cite{Lucas2018}, which is calculated by dividing the consumed energy by the rated energy (i.e. energy that would be consumed by the charging pool if working at the maximum capacity for the whole observed period). This indicator may inform power grid operators about the possible supply margin that could be eventually used by other consumers and it is one of the indicators that can help charging infrastructure operators to understand better the utilization of charging pools.
\end{itemize}

\begin{table}[ht]
	\centering
	\begin{tabular}{ll}
	\hline
        Charging pool performance indicator & $R^2$ \\
	\hline
        Consumed energy [kWh] & 0.44\\
		 Number of charging transactions & 0.50\\
		 Popularity (unique number of RFID cards) & 0.60\\
		 Charging time [hours] & 0.48 \\
		 Charging ratio (charging time divided by connection time) & 0.40 \\
		 Use-time ratio (charging time divided by overall operation time) & 0.42 \\
		 Energy ratio (consumed energy divided by rated energy) & 0.38 \\
	\hline
	\end{tabular}
	\caption{Assessment of performance indicators using predictors characterizing urban context and human activities near charging pools. The values of all performance indicators were calculated from the 2015 data. The coefficient of determination, $R^2$, was obtained by the ordinary least squares method.} 
	\label{tab:R2regResponses}
\end{table}

The values of all performance indicators were calculated from the 2015 data. From the EVnetNL data, we found that if a charging pool has more than one connector, connectors were used only very rarely simultaneously. In the case of a charging pool with two connectors, this is because of construction reasons, in other cases this is probably due to a still relatively low number of electric vehicles in 2015. For this reason, the proposed indicators measure the performance of charging pools without discounting for the number of connectors. To analyze how well predictors derived from GIS data fit proposed performance indicators, we applied the ordinary least squares methods to each performance indicator separately. In Table~\ref{tab:R2regResponses}, we report values of the coefficient of determination, $R^2$. Results show that the highest $R^2$ value and hence the highest potential for data analysis is found for the popularity of charging pools (expressed by the unique number of RFID cards used to initiate charging). Therefore we limit further analyses presented in this paper to this performance indicator. To ensure transparency of presented analyses, as a supplementary material we publish together with the paper files containing values of investigated performance indicator and the matrix of predictors.

When planning the deployment of new infrastructure, often we need to select from a finite number of candidate sites (locations where it is feasible to install a charging infrastructure from the perspective of land ownership, supply with energy, potential to attract sufficient demand, etc.). In such a situation, it is not necessary to estimate the exact number of EV drivers attracted by a charging pool but to provide a ranking of candidate sites. For this reason, we reduce the problem that we address in this paper, to the problem of predicting whether a given candidate site belongs to the top rank candidate sites or not. This problem can be formalized as a classification problem. In the next section, we briefly introduce selected classification methods, logistic regression with an $\mathit{l}_1$ penalty, gradient boosted regression trees and random forests. 

\section{Methods}
\label{sec:methods}
To describe classification methods, first we introduce a basic notation. We denote matrix of predictors as $\bold{X} \in \mathbb{R}^{n \times p}$. The matrix $\textbf{X}$ is formed by $i = 1, \dots, n$ observations $\textbf{x}_i = \{x_{i1}, \dots, x_{ip} \}$ (rows of $\textbf{X}$) and $j = 1, \dots, p$ predictor vectors $x_j~=~\{x_{1j}, \dots, x_{nj} \}^T$ (columns of $\textbf{X}$). A vector of response variables, we denote as $y = \{y_1, \dots, y_n\}$, with $y_i \in \{0, 1\}$ for $i = 1, \dots, n$, where $y_i = 1$ represents the situation in which the $i$-th charging pool belongs to the top $z\%$ of most popular stations and $y_i = 0$ otherwise.

Finally, from collected data we derived $p = 172$ predictors, each with $n=1271$ observations (charging pools). Although, $n > p$ holds, considering the number of observations, the number of predictors is relatively high. We assume that only a relatively small fraction of predictors plays an important role, i.e. we expect that the resulting model will be sparse. For this reason, we selected classification methods that can handle well sparsity~\cite{Hastie_2009,Hastie_2015}. 

\subsection{Logistic regression with $\mathit{l}_1$ penalty}
\label{sec:l1_logistic_regression}
The logistic regression is a popular method for binary classification problems leading to solving a convex optimization problem~\cite{Boyd_2004}. The response is modeled as a random variable $Y \in\{0,1\}$ and the observation is modeled as a random variable $X \in \mathbb{R}^{p}$. The logistic model takes the form

\begin{equation}
P(Y = 1|X = x) = \frac{1}{1+e^{-(\beta_0 + \beta^T x)}},
\label{eq:log_reg}
\end{equation}
where $\beta_0$ is the intercept and $\beta$ is the vector of regression coefficients. The maximum likelihood estimate of parameters $\beta_0$ and $\beta$ in Eq.~(\ref{eq:log_reg}) is found by solving the optimization problem

\begin{align}
\underset{\beta_0, \beta}{\text{maximize }} & \frac{1}{n}\sum_{i=1}^{n}\left \{{y_i(\beta_0 + \beta^{T} \textbf{x}_i)-\log{(1+e^{\beta_0 + \beta^T \textbf{x}_i})}}\right\}.
\label{eq:log_reg_obj}
\end{align}

Likewise, as in the LASSO method~\cite{Tibshirani_1996}, the logistic regression with $\mathit{l}_1$ penalty (LR-$\mathit{l}_1$) is obtained by adding $\mathit{l}_1$ regularization to the objective~(\ref{eq:log_reg_obj}), resulting in the optimization problem

\begin{align}
\underset{\beta_0, \beta}{\text{maximize }} & \frac{1}{n}\sum_{i=1}^{n}\left \{{y_i(\beta_0 + \beta^{T} \textbf{x}_i)-\log{(1+e^{\beta_0 + \beta^T \textbf{x}_i})}}\right\} + \lambda \left\|{\beta}\right\|_1,
\label{eq:log_reg_lasso_obj}
\end{align}
that is solved for some $\lambda \geq 0$. The $\mathit{l}_1$ penalty enables to shrink less-informative coefficients $\beta$ to zero and thereby increases the simplicity and explanatory power of the model. Hyperparameter $\lambda$ in the objective function~(\ref{eq:log_reg_lasso_obj}) enables to set a trade-off between the quality of the fit and sparsity of the model. 

Equation~(\ref{eq:log_reg}) is also used to derive predictions. Estimated values of regression coefficients $\hat{\beta}_0$ and $\hat{\beta}$ together with the observation $\textbf{x}$ are plugged into the right hand side of Eq.~(\ref{eq:log_reg}) and the resulting value is used as an estimate $\hat{y}$. Using hyperparameter $\theta$, the thresholding is used to transform $\hat{y}$ to the binary value. Hence, if $\hat{y} \geq \theta$ the prediction is $1$ and otherwise it is $0$.

\subsection{Random forests}
Random Forests (RF) is a method based on the regression tree model~\cite{Hastie_2009}. The regression tree model predicts the target variables from ramified observations. More specifically, this method splits the training data into several subsets by applying conditions upon predictors. The model training represents a ramification process. When the tree is trained, branches grow from a single node, and every node determines a condition on a single predictor. A unique path is traced on the basis of the value of a single predictor, iteratively splitting the datasets into two children subsets. In order to determine the local optimal condition for the split, the Mean Squared Error (MSE) is minimized. RF allows diversifying the training of multiple regression trees, \cite{Breiman2001}. Particularly, a number \textit{m} of individual trees (i.e. the RF) is independently trained using a bagged (bootstrap aggregated) subset of the total training data.
The \textit{m-}th regression tree generates a prediction through the following equation:
\begin{equation}
\hat{y}^{(m)} =\sum_{i=1}^{n} w_i^{(m)}(\textbf{x})\cdot y_i,
\label{eq:regrtree}
\end{equation}
where $\textbf{x}$ is an observation and ${w_i^{(m)}}(\textbf{x})$ is weight evaluated as follows:

\begin{equation}
w_i^{(m)}(\textbf{x})=\frac{1\{\textbf{x}_i\in R_{L^{(m)}}\}}{\sum_{j=1}^{n}1\{\textbf{x}_j\in R_{L^{(m)}}\}},
\label{eq:RFweights}
\end{equation}
with $L^{(m)}$ the leaf of the \textit{m-}th tree individuated by $\textbf{x}$ , and $R_{L^{(m)}}$ the domain of this leaf. The function $1(\cdot)$ takes value 1 if the expression within the brackets is true and 0 otherwise. Thus, the \textit{m}-th tree returns as a prediction the average value of all responses that belong to the same leaf node as the observation for which the prediction is made. Then, RF returns its prediction through a simple average of the predictions of the individual trees. Finally, to obtain binary prediction, thresholding is applied.

\subsection{Gradient boosting regression tree}
The Gradient boosting regression tree (GBRT) exploits regression trees, within a different framework. In this case, regression trees are fitted upon residuals of a weak learner in an iterative way. The fitting stops when the improvement brought by the last iteration is smaller than a fixed threshold~\cite{friedman2001}. 
At the generic iteration \textit{m}, the prediction is given by a recursive equation:
\begin{equation}
F_m(\textbf{x})=F_{m-1}(\textbf{x})+\sum_{j=1}^{J^{(m)}}\gamma_j^{(m)} \cdot 1\{\textbf{x}\in R_{j^{(m)}}\},
\label{eq:GBoutcome}
\end{equation}
where \textit{F} is the prediction provided by the weak learner, $J^{(m)}$ is the number of terminal regions $R_1^{(m)}, \dots, R_{J^{(m)}}^{(m)}$ (each corresponding to one leaf node) and $\gamma_1^{(m)},\dots , \gamma_{J^{(m)}}^{(m)}$ are parameters estimated at each iteration \textit{m}. The estimation of these parameters is accomplished by minimizing the MSE. The equation (\ref{eq:GBoutcome}) gives the prediction of the target variable. Finally, to obtain binary prediction, thresholding is applied.

\section{Results}
\label{sec:results}
In this section, the predictability of the popularity of charging pools is evaluated using various metrics. Since the LR-$\mathit{l}_1$ classification method is returning a sparse vector of regression coefficients, we evaluate the type and strength of the influence of predictors on the popularity as well. 

\subsection{Measures of predictability }
A set of measures (see Table~\ref{tab:measureEquations}) was compiled to assess the performance of classification models from different perspectives. All measures can be calculated from the elements TP (true positives), TN (true negatives), FP (false positives) and FN (false negatives) of the confusion matrix~\cite[p.254]{kuhn2013applied}.

\begin{table}[ht]
	\centering
	\setlength{\extrarowheight}{7pt}
	\begin{tabular}{lc}
		\hline
		\makecell{Predictability \\ measure} & \makecell{Mathematical \\ expression}\\ 
		\hline
		\textit{Accuracy} & $\frac{TP + TN}{n}$ \\ 
		\textit{Precision} & $\frac{TP}{TP + FP}$ \\ 
		\textit{Sensitivity} &  $\frac{TP}{TP + FN}$ \\ 
		\textit{Fall--out} & $\frac{FP}{FP + TN}$ \\ 
		\textit{F--score}  & $2\frac{Sensitivity\cdot Precision}{Sensitivity + Precision}$ \\
		\makecell{\textit{Matthews’ correlation} \\ \textit{coefficient} (MCC)}& $\frac{TP\cdot TN-FP\cdot FN}{\sqrt{(TP+FP)(TP+FN)(TN+FP)(TN+FN)}}$\\
		\hline
	\end{tabular}
	\caption{Overview of predictability measures and mathematical descriptions showing how the predictability measures are calculated from the elements of the confusion matrix.} 
	\label{tab:measureEquations}
\end{table}

The \textit{accuracy} is the proportion of pools predicted correctly, the \textit{precision} is proportion of correct predictions of popular pools and the \textit{sensitivity} (also called true positive rate) is the proportion of popular pools predicted correctly. The \textit{fall--out} (called also false positive rate) is a fraction of unpopular pools predicted incorrectly and it is used on the x-axis in the receiver operating characteristic (ROC) curve. The overall performance of a classifier can be evaluated by the area under the ROC curve (AUC) \cite[p.~147]{james2013introduction}.

By analyzing the dependency between a measure of predictability and probabilistic threshold $\theta$ a suitable range for $\theta$ can be determined. When applying this approach to the accuracy for data with class imbalance (i.e. unequal number of $1$s and $0$s in the response vector) it may not be intuitive to determine models reaching good accuracy (e.g. for $25\%$ of $1$s in the response vectors, we reach value of accuracy $0.625$ by just randomly guessing $1$s with probability of $0.25$ and $0$s with probability of $0.75$). The F--score and Matthews’ correlation coefficient (MCC) are more balanced measures recommended for data with class imbalance~\cite{kuhn2013applied,matthews1975comparison}. The \textit{F--score} combines precision and sensitivity, taking harmonic mean of both measures, i.e. it moves towards the lower of the values~\cite{sasaki2007truth}. By definition, if any of the values in the parentheses in the denominator of the equation defining MCC is $0$ (see Table~\ref{tab:measureEquations}), the MCC is set to $0$. For models predicting better (worse) than a random model, the MCC is positive (negative). Models as skilled as a random guess have MCC equal to $0$. The MCC equals to $1$ ($-1$) if all the observations are predicted correctly (wrongly).

\subsection{Settings of parameters} 
A radius of the buffer, representing vicinity of a pool, was chosen from the set $\{100, 150, 200, 250, 300, 350, 400, 450, 500\}$ meters. This range is the distance drivers typically walk from the parking place to their destination~\cite{Waerden_2015}. The dependency between the popularity of charging pools and predictors was fitted by the least squares model for all values in the set. The highest value of $R^2$ was found for the radius of $350$ meters and selected for further analyses. In experiments, training and testing sets are assigned $80\%$ and $20\%$, respectively, using stratified sampling with respect to the response variable~\cite[p.68]{kuhn2013applied}. To gain more reliable results and conclusions, we evaluate variability of calculated measures by analyzing outputs of 100 models, trained on 100 different splits into training and testing sets. While evaluating predictability measures, we varied threshold $\theta$ in the range from 0 to 0.99, in steps of 0.01. The hyperparameter $\lambda$ in Eq.~(\ref{eq:log_reg_lasso_obj}), was found by k-fold cross validation, where the stratified sampling was used to split data into $k=10$ folds preserving the class distribution of the response variable. Considering values $10^{-4 + i*0.015}$, for $i \in\{0, ..., 200\}$, $\lambda$ is assigned the value corresponding to the largest AUC value~\cite{friedman2010regularization}. Similarly, when growing decision trees, the k-fold cross validation was applied to set the number of learning cycles, the learn rate for shrinkage, the minimum size of leaves, and the maximum number of splits. 
The values of predictability measures were obtained by applying the trained model to the testing dataset. The response vector was encoded into a binary format by setting $25\%$ of the elements corresponding to top charging pools ranked by the popularity to value $1$ and the rest to $0$. We did also experiments with values of $15\%$, $20\%$, $30\%$ and $35\%$, however, very similar conclusions could be drawn from the results and we do not report them. In computations, we used $\mathit{l}_1$-regularized logistic regression implemented by the R package \textit{glmnet}~\cite{friedman2010regularization}. GBRT and RF were implemented in MATLAB environment within Statistic and Machine Learning Toolbox. 

\subsection{Predictability of popularity of charging pools}
\begin{figure}
	\centering
	\includegraphics[width=0.97\textwidth]{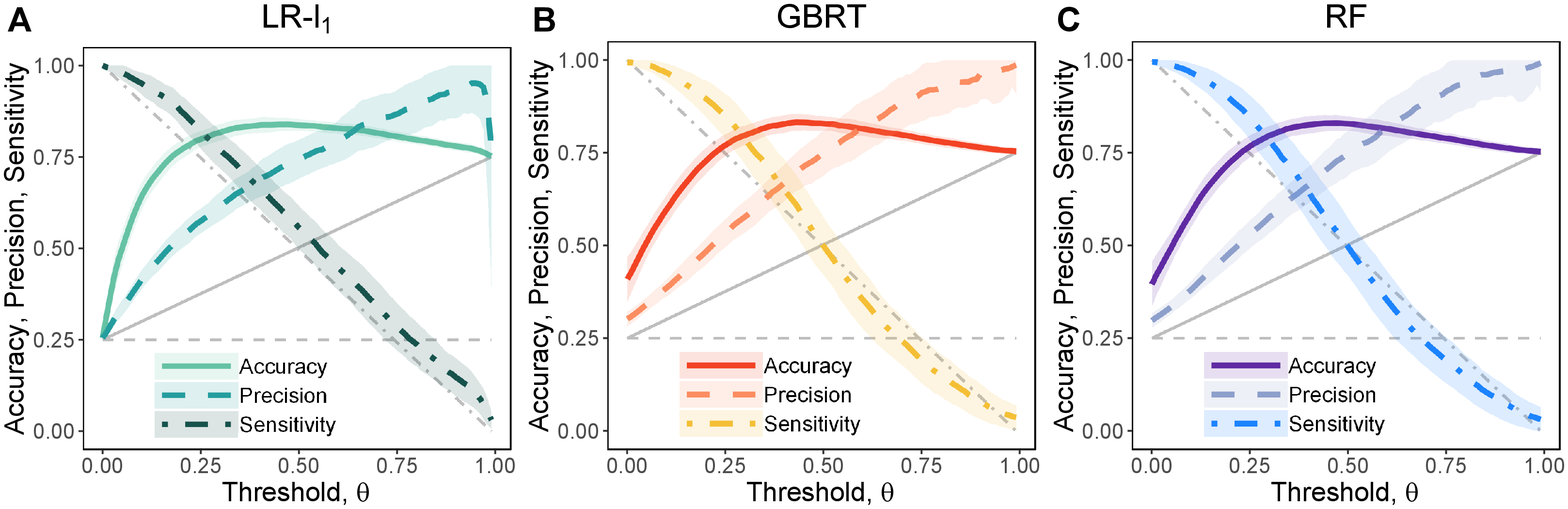}
	\caption{The mean value of the accuracy, precision and sensitivity evaluated on testing data as a function of threshold $\theta$ in (A) for LR-$l_1$, in (B) for GBRT and in (C) for RF method. Each measure is displayed in different line style. Thick lines show average values over an ensemble of 100 different training and testing datasets splits and shaded areas represent one standard deviation. Thin grey lines show the expected value of measures, achieved by the random model predicting popular pools with probability of $0.25$.} 
	\label{fig:acc_prec_sens}
\end{figure}

\begin{figure}
	\centering	
	\includegraphics[width=0.97\textwidth]{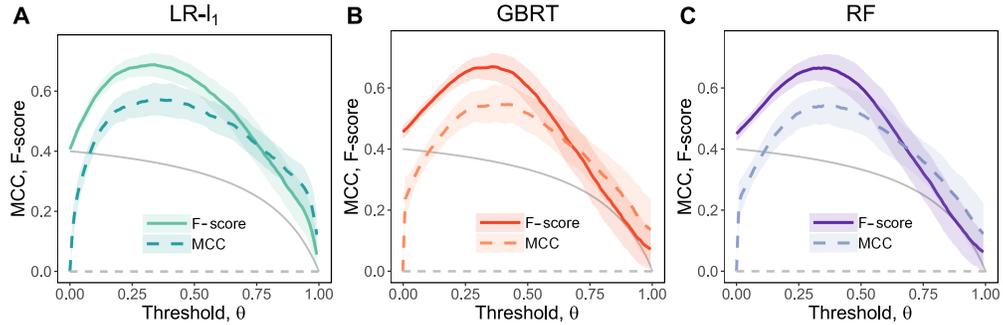}
	\caption{Mean value of the F--score and MCC measures evaluated on testing data as a function of threshold $\theta$ in (A) for LR-$l_1$, in (B) for GBRT and in (C) for RF method. Lines show average values over an ensemble of 100 different training and testing datasets splits and shaded areas represent one standard deviation.Thin grey lines show the expected value of measures, achieved by the random model predicting popular pools with probability of $0.25$.} 
	\label{fig:MCC_YJS}
\end{figure}

In Figure~\ref{fig:acc_prec_sens}, the mean accuracy, precision and sensitivity are shown for all three methods. To facilitate evaluation of the quality of predictions, we consider a null model predicting popular stations randomly with probability of $0.25$. It can be easily shown that considered measures for the null model take the functional forms presented in Table~\ref{tab:nullModelEquations} and displayed in Figures~\ref{fig:acc_prec_sens} and~\ref{fig:MCC_YJS} by thin lines.

\begin{table}[ht]
	\centering
	\setlength{\extrarowheight}{7pt}
	\begin{tabular}{lc}
		\hline
		Measure & Function \\ 
		\hline
		\textit{Accuracy} & $0.25 + 0.5 \theta $ \\ 
		\textit{Precision} & $1 - \theta $ \\ 
		\textit{Sensitivity} &  $0.25$ \\ 
		\textit{F--score} & $\frac{1-\theta}{2.5-2\theta}$ \\
		\textit{MCC} & $0$ \\
		\hline
	\end{tabular}
	\caption{Functional forms describing the expected value of evaluated measures for the considered null model as a function of the threshold~$\theta$.} 
	\label{tab:nullModelEquations}
\end{table}

All three methods outperform the null model, in accuracy and precision measures in the whole range of $\theta$. It is unavoidable that the sensitivity is decreasing as the threshold $\theta$ grows. For $\theta$ values greater than $0.5$ the sensitivity falls below 0.5, which results in low applicability of predictions. As $\theta$ grows, the sensitivity can reach value zero, even for an ideal model, if the threshold $\theta$ is already too high. If $\theta$ is larger than 0.5, for the decision tree methods we find the sensitivity to be smaller than the null model confirming that such thresholds are too high.

In the literature, various approaches on how to select the value of the threshold $\theta$ and hence to find a reasonable trade-off between accuracy, precision and sensitivity can be found~\cite{kuhn2013applied}. We selected two metrics, MCC and F--score, which are evaluated in Figure~\ref{fig:MCC_YJS}. Both measures reach one single maximum, which is hence also the global maximum. Values of the threshold where the maxima are achieved we denote as $\theta_{MCC_{max}}$ and $\theta_{F\textendash score_{max}}$. Values of the accuracy, precision and sensitivity for $\theta_{MCC_{max}}$ and $\theta_{F\textendash score_{max}}$ are reported in Table~\ref{tab:thetaResults}. 

\begin{table}[ht]
	\centering
	\begin{tabular}{r|rrr}
		\hline
		& LR-$l_1$ & GBRT & RF \\
		\hline
		$\theta_{MCC_{max}}$ & 0.34 & 0.35 & 0.37 \\ 
		\hline
		Accuracy & 0.829  & 0.823 & 0.819 \\
		Precision & 0.648 & 0.644 & 0.632 \\
		Sensitivity & 0.728 & 0.694 & 0.701 \\
		MCC & 0.571 & 0.548 & 0.542 \\
		\hline
		$\theta_{F\textendash score_{max}}$ & 0.34 & 0.35 & 0.35 \\
		\hline
		Accuracy & 0.829  & 0.813 & 0.813 \\
		Precision & 0.633 & 0.614 & 0.615 \\
		Sensitivity & 0.750 & 0.736 & 0.727 \\
		F--score & 0.685 & 0.667 & 0.664 \\
		\hline
	\end{tabular}
	\caption{The mean values of the accuracy, precision and sensitivity from Figure~\ref{fig:acc_prec_sens} for all three methods and selected threshold values $\theta_{MCC_{max}}$ (threshold value corresponding to the maximum value of the MCC measure) and $\theta_{F\textendash score_{max}}$ (threshold value corresponding to the maximum value of the $F\textendash score$ measure).} 
	\label{tab:thetaResults}
\end{table}

When decisions about the extension of the existing charging infrastructure are taken, many often contradicting factors are taken into account. Alternatively, stakeholders can select a threshold $\theta$ based on their expectations and attitudes towards risk. The lower the $\theta$, the more likely is identification of popular locations with the drawback of increased risk of placing charging pools into unpopular areas. In opposite, the higher $\theta$, we identify the popular charging pools with higher assurance, with the drawback of overlooking potentially popular locations. 
According to observed values of measures, we recommend considering the threshold $\theta$ within the range from 0.3 to 0.45, where both precision and sensitivity are relatively high.

Comparison with the null model and values of measures in Table~\ref{tab:thetaResults} indicate that urban area and characteristics of charging pools contain some predictive power for the popularity. What values of measure justify good quality of models typically depend on the application domain~\cite[p.70]{james2013introduction}. Considering data analyses concerning the human choice in similar domains such as for example bike sharing applications~\cite{zhang2016bicycle, chen2017understanding}, the obtained values of the accuracy exceeding value 0.8, while both precision and sensitivity are larger than 0.65, can be considered as favorable. 

\begin{figure}
	\centering
	\includegraphics[width=0.97\textwidth]{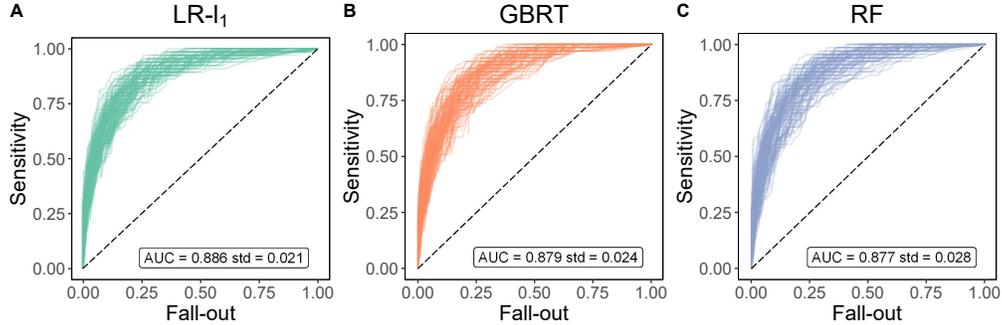}
	\caption{Ensemble of 100 ROC curves (sensitivity versus fall-out) corresponding to different training and testing datasets splits. The dashed line visualizes points corresponding to the random classifier. Value of the area under the curve (AUC) and the corresponding standard deviation are reported in the frame located at the bottom.} 
	\label{fig:ROC}
\end{figure}

Simple inspection of Figures~\ref{fig:acc_prec_sens}-~\ref{fig:MCC_YJS} indicates, that all three methods provide similar results. To evaluate, whether the results are statistically distinguishable, we test the differences in AUC ensembles, i.e. areas under the ROC curves. The ROC curves corresponding to 100 different training and testing dataset splits are shown in Figure~\ref{fig:ROC}. In all statistical tests we set the significance level $\alpha = 0.01$. 

In the first step, the equality of AUC variances between all pairs of methods was tested using the F-test. Statistically significant difference was identified only for LR-$\mathit{l}_1$ and RF methods, the first having less variance ($p = 0.0017$). To compare mean values of AUCs, for the case when the variances were not found to be significantly different, the t-test was used otherwise, the Welch t-test \cite{welch1947generalization}. The mean AUC of LR-$\mathit{l}_1$ was larger than RF ($p = 0.0094$), and not different from GBRT ($p = 0.0379$). The mean AUCs of RF and GBRT were not significantly different ($p = 0.5091$). Hence, according to AUC values, the method LR-$\mathit{l}_1$ is more stable than RF, and outperforms the method in mean values of AUC. On the significance level $\alpha = 0.05$, LR-$\mathit{l}_1$ method has a significantly larger mean value of AUCs than both, the GBRT and RF methods.

\subsection{Characteristics influencing the popularity of public charging stations}
The shrinkage nature of $\mathit{l}_1$ penalty in the LR-$\mathit{l}_1$ method provides better predictions on testing data and variable selection functionality. This gives an advantage compared to decision tree methods, that yielded clumsy models, involving 169 predictors on average, making them hard to interpret. Due to this reason, we interpret the results for the LR-$\mathit{l}_1$ method only.

The values $\hat{\beta}$ of the LR-$\mathit{l}_1$ model might be sensitive to the used sample of training data, hence an analysis of the results robustness is necessary. The conventional solution is to evaluate the $p$-values of coefficients estimated by statistical methods. The problem to calculate $p$-values for LR-$\mathit{l}_1$ model is difficult due to the adaptive nature of the estimation procedure~\cite{tibshirani2015statistical}. Therefore, to capture the stochasticity in coefficients $\hat{\beta}$, we estimate their distributions by sampling the dataset 500 times using bootstrap~\cite[p.~187]{james2013introduction}. A model is fitted to each bootstrapped dataset using stratified cross-validation. To facilitate a comparison of the impact of predictors, we standardize each element of $\hat{\beta}$ by dividing it with the sample standard deviation of the corresponding predictor.

\begin{figure}
	\centering
	\includegraphics[width=0.97\textwidth]{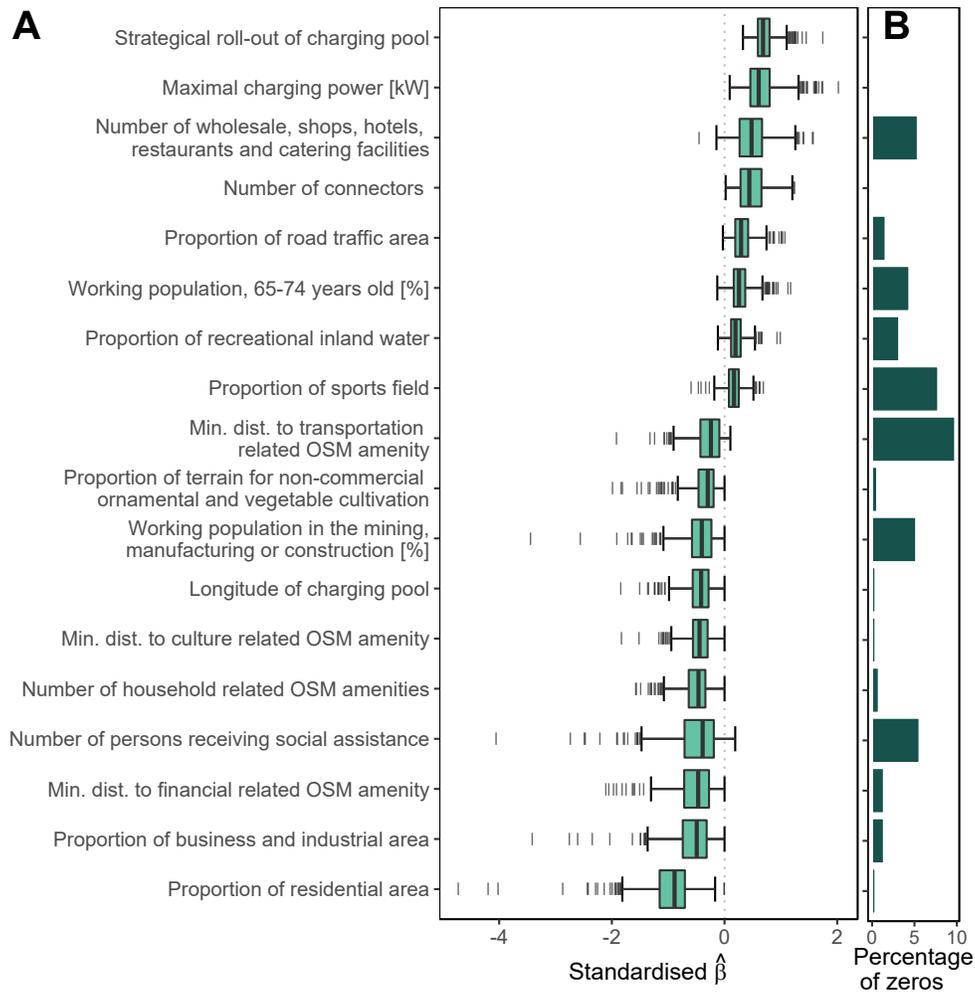}
	\caption{
		(A) Tukey Box-plot of standardized coefficients $\hat{\beta_j}$ for 500 $LR-l1$ models. The coefficients are sorted by median values. Only predictors that were selected by at least 90\% of models are displayed. (B) The percentage of models setting the standardized coefficient $\hat{\beta_j}$ to zero.
	} 
	\label{fig:all_var_importance} 
\end{figure}

The group of selected predictors depends on the training data sample. In Figure~\ref{fig:all_var_importance}, sampled distributions of coefficients that have been selected most frequently (at least by 90\% out of 500 models) are displayed. The bar plot presents the percentage of models where the coefficients were equal to zero.

The impact of a predictor on the response increases with the absolute value of the corresponding coefficient~\cite{kuhn2013applied}. Positive (negative) sign of a coefficient indicates increasing (decreasing) impact of the predictor. A way how to quantify the significance of coefficient $\hat{\beta}$ is to assess the likelihood that the coefficient is different from zero. Analysis of the distribution of coefficients $\hat{\beta}$ allows us to make such assessments and to conclude how certain is the positive or negative influence of predictors on the response variable.
 
In summary, selected predictors can be categorized into three groups: the function of the geographic area constituting the vicinity of a charging pool, characteristics of the population living in this area and properties of charging pools. From the perspective of the geographic area, the most important predictors are the number of wholesale businesses, shops, hotels, restaurants and catering businesses, areas with recreational inland water, sports fields and roads, all having a positive impact. The minimal distance to financial, cultural, and transportation OpenStreetMap (OSM) amenities have negative coefficients, meaning that large minimal distance is decreasing the popularity of charging pools. Hence, if these facilities are found in the proximity of charging pools, they have a positive impact on the popularity. 

In opposite, the residential areas, the areas with non-commercial ornamental and vegetable cultivation and the presence of OSM amenities related to households tend to lower the popularity of charging pools. These findings are well aligned with the intuition that in residential areas the charging pools are visited by more homogeneous groups of users than in the more crowded urban areas. Similarly, the most likely explanation of the negative impact of business and industrial areas is work charging~\cite{Sadeghianpourhamami_2018}, i.e. either charging of a fleet of company cars or regular use of charging pools by (a small group of) employees commuting to work.  

A notable population group living near popular charging pools is working elderly people between 65 and 74 years old. In opposite, popularity is negatively correlated with areas inhabited by the population working in the mining, manufacturing or construction sector as well as persons who depend on social assistance. Thus, these results suggest that the economic prosperity of the population in the vicinity is affecting the visiting patterns of charging pools. The popular charging pools are more likely to be deployed following strategical rollout and have larger maximal power and more connectors. The negative influence of geographic longitude can be explained by the geography of the Netherlands, while the western part of the country is more urbanized and we can find here the majority of large Dutch cities.

\section{Discussion and conclusions}
\label{sec:discussion_and_conslusions}
This study demonstrates the ability of classification methods to predict popular locations of charging pools from various GIS and large-scale charging infrastructure data. Moreover, we evaluate the impact and significance of factors affecting the popularity of charging infrastructure. Predicting the popularity of charging pools is of utmost importance for matching EV requirements driven by social habits with energy requirements related to electrical network configuration. Main conclusions derived from the data analysis are the following:
\begin{itemize}
	\item 
	To characterize the performance of charging infrastructure is a complex task as various viewpoints need to be considered. We selected seven indicators characterizing performance considering energy supply issues, charging demand and expectations of infrastructure operators. The popularity of charging pools represented by the number of unique RFID cards can be explained to the largest degree among the seven indicators of charging infrastructure.
	
	\item To exploit the previous finding, we formulate the classification problem of determining top charging pools, ranked by the popularity. The $\mathit{l}_1$-regularized logistic regression, gradient boosted decision trees and random forests, are able to predict the popularity with the accuracy exceeding value 0.8 and F--score reaching value 0.68, clearly outperforming random models. Such values do not justify decisions taken solely based on predictions provided by created models, however, results from our models can give indications for decision making processes in charging infrastructure planning.
	
	\item Factors having positive influence on the popularity of charging pools are rollout strategy, maximal power, number of connectors, number of shopping, catering, sport and trade related venues. Hence, charging pools located on frequently visited spots and providing convenient charging opportunities are more likely to become popular. The largest negative influence have residential, business and industrial areas and areas inhabited by workers and persons receiving social benefits. Thus, areas inhabited by social groups with lower purchase power and areas periodically visited by a small group of EV drivers are associated with lower popularity.	
\end{itemize}

The presented results are limited by the low utilization of some charging pools, which can be attributed to the low penetration of electric vehicles in some areas in the Netherlands. Thanks to the rapid growth of EVs, the utilization of charging pools is expected to increase and potentially mitigate this limitation. A difficulty often present in GIS data is the interdependence of factors, expressed as collinearity, causing nontrivial problems when interpreting the impact of individual factors. There is no generally accepted approach to address this problem. We minimize the chances that collinearity affects the results by removing highly correlated factors. Nevertheless, predictors that we present as influential and significant, should be taken into account with some care. Typically, the uncaptured stochasticity of models can be attributed to missing data. We assume that more detailed mobility data, e.g. GPS and floating car data could improve the results. Geographically, our study is focused on the area of the Netherlands, which might impose some limitations when transferring models and conclusions to other countries. Although, we expect similar results for comparable geographic and demographic contexts.

This study opens several directions. For instance, future research could explore possibilities how to design prediction models for other performance indicators of charging pools, how to efficiently downscale prediction models to a level of a region or a city or how to improve predictions by customizing models to specific classes of charging pools. Another challenge is the application of regression approaches that could successfully predict the values of performance indicators.

\section*{Acknowledgements}

This work was supported by the research grants: VEGA 1/0089/19 ”Data analysis methods and decisions support tools for service systems supporting electric vehicles”, VEGA 1/342/18 ”Optimal dimensioning of service systems”, APVV-15-0179 ”Reliability of emergency systems on infrastructure with uncertain functionality of critical elements”, Operational Program Research and Innovation in frame of the project: ICTproducts for intelligent systems communication, code ITMS2014+ 313011T413, co-financed by the European Regional Development Fund and by the Slovak Research and Development Agency under the contract no. SK-IL-RD-18-005. We thank \v{L}udmila J\'{a}no\v{s}\'{i}kov\'{a} and Luca Lena Jansen for valuable comments and suggestions.

\bibliographystyle{cas-model2-names}

\section*{Author contributions statement}
CRediT (Contributor Roles Taxonomy) has been applied to describe contributions of authors. \textbf{Milan Straka:} Data curation, Methodology, Formal analysis, Investigation, Validation, Visualization, Software, Writing-original draft; \textbf{Pasquale De Falco:} Formal analysis, Methodology, Investigation, Software, Validation, Writing-original draft; \textbf{Gabriela Ferruzzi:} Validation; Writing original draft; Writing-review and editing; \textbf{Daniela Proto:} Validation; Writing original draft; Writing-review and editing; \textbf{Gijs van der Poel:} Conceptualization, Resources, Writing-review and editing; \textbf{Shahab Khormali:} Resources, Writing-original draft, Writing-review and editing;
\textbf{\v{L}ubo\v{s} Buzna:} Conceptualization, Funding acquisition, Methodology, Resources, Supervision, Writing-review and editing.

\section*{Additional information}
Competing financial interests: The authors declare no competing financial interests.

\newpage
\setcounter{section}{0}
\section*{Supplementary Information}

The supplementary information file contains an extended description of datasets, together with descriptions of the predictor extraction techniques and data pre-processing procedures. The final section contains a table summarizing predictors that are published together with the paper.

\section{EVnetNL dataset processing}
\label{secSI:EVnetNL_dataset}

The EVnetNL dataset was provided for research purposes by the Dutch innovation company ElaadNL~\cite{elaadnl}. This dataset contains more than 1 million charging transactions of electric vehicles that are spread over more than 4 years taking place from January 2012 till March 2016. Each charging transaction is described by the time of plugging in, time of plugging out, amount of transferred energy in kWh and the length of the time interval that was used to transfer energy from the charger to the vehicle (charging time). The time interval, when a vehicle is plugged in, but no energy is transferred from the charger to the vehicle, we call idle time. Charging transactions are associated with EV drivers via the RFID card identifier and with charging stations that are identified by a unique name and GPS coordinates. The dataset contains 1~747 unique names of charging stations that are distributed across the whole area of the Netherlands, where the vast majority of stations have either one or two connectors. Each charging connector is characterized by the maximum charging power in kW, where all values included in the dataset are below 11~kW meaning that all connectors are slow chargers. Each charging station is associated with the type, taking one of the two following values: strategic or demand-driven. The type of the station characterizes the used rollout strategy when deploying stations, while demand-driven stations were placed upon requests from EV drivers, strategic stations were located to cover uniformly the area of the Netherlands.

Incomplete or inconsistent charging transaction data were discarded. For the analyses, we filtered only charging transactions that were initiated after  December 31, 2014, and terminated before January 01, 2016. Hence, we used the last available complete year in the dataset when the number of charging stations was already relatively stable~\cite{Flammini_2019}. We aim to characterize the vicinity of charging stations with available GIS data. Hence, if two charging stations are very close to each other, we cannot properly distinguish them in the analysis. For this reason, we adopted the concept of charging pools derived from data. If the Euclidean distance between two charging stations was less than 50 meters, we aggregated them and formed a charging pool. After applying these steps, we obtained data that contains  373~302 charging transactions, 32~074 unique RFID cards, 1~271 charging pools that are equipped with 2~155 connectors. Among the charging pools, 749 were strategically located and 522 were demand-driven located.

\section{GIS datasets}
\label{secSI:GIS_datasets}
Various open GIS datasets were collected, to model population, urbanity and businesses in the vicinity of the charging pools. For the overview, see Table~\ref{tabSI:GIS_data}. Two datasets, Population cores, and Neighborhoods directly represent population characteristics by using spatial statistical units of neighborhoods and population cores, respectively. The population is described from the perspective of age categories, income groups, household composition, etc. The LandScan dataset provides 24-hour average population density estimate in a raster format. The Liveability dataset describes the quality of living in the resolution of neighborhoods, using 1 general and 5 specific indexes regarding housing, residents, services, safety, and living environment. The Energy dataset contains gas and electricity consumption data of households and companies in the resolution of neighborhoods. The boundaries of land use areas are based on the digital topographical base Top10NL. Land use data associate one out of twenty-five available land use categories to each spatial unit. Examples of land use categories are roads, residential areas, business areas, socio-cultural facilities, public facilities, etc. Traffic flows dataset contains estimates of count statistics for 3 vehicle categories (cars, vans, trucks) and 3 periods of the day (7--19hr, 19--23hr, 23--7hr) in high resolution of individual road segments. From the OpenStreetMap dataset, points of interest (POI) organized in 593 categories were extracted to capture locations of venues that EV drivers might be interested to visit. As some of the POI categories contain very sparse data, we aggregated them into 15 categories (e.g. health, entertainment, finance, fashion, food, transportation, education, sport, family, etc.) based on shared characteristics. Charging pools 2015 is a set of publicly available charging pool locations in the Netherlands in the year 2015. This dataset was compiled from EVnetNL, OpenChargeMap~\cite{ocm} and OplaadPalen~\cite{opp}. First, we included all charging pools from the EVnetNL dataset to the Charging pools 2015  dataset. Second, we processed one-by-one all charging stations from OpenChargeMap and OplaadPalen (in this order) and we added to the set of charging pools those charging stations that were distant 50 meters and more from already included charging pools. Whether a charging pool derived from OpenChargeMap and OplaadPalen was already existing during the year 2015, we estimated from the date(s) when the corresponding station(s) was(were) included to databases.

\begin{sidewaystable}
	\begin{longtable}[h]{p{.2\textwidth} p{.35\textwidth} p{.15\textwidth} p{.15\textwidth}  } 
		\hline
		Dataset &  File format/resolution & Attributes & Predictors\\
		\hline
		Population cores & shapefile/heterogeneous polygons & 104 & 46 \\
		Neighbourhoods & shapefile/neighbourhoods & 120 & 64 \\
		Land use & shapefile/custom & 1 & 25 \\ 
		Energy & shapefile/neighbourhoods & 12 & 12 \\
		Liveability & shapefile/neighbourhoods & 6 & 6 \\
		Traffic flows & shapefile/road segments & 9 & 7 \\
		LandScan & raster/30$\times$30 arc seconds & 1 & 1 \\ 
		OpenStreetMap & point data & 15 & 30 \\
		Charging pools 2015 & point data & 1 & 2 \\
		\hline
		\caption{Overview of publicly available GIS databases for the territory of the Netherlands that have been identified as relevant for the analyses. The column  "Attributes", reports the number of data attributes associated with each geometric object (point, polyline, polygon) included in the database. "Predictors" column stands for the number of derived predictors from the data attributes, after the data pre-processing.}
		\label{tabSI:GIS_data}
	\end{longtable}
\end{sidewaystable}

\section{Data pre-processing}
\label{secSI:data_pre_processing}

\subsection{Missing values handling}
\label{secSI:missing_values_handling}

Preliminary analysis of GIS attributes revealed some cases of missing data mainly occurring in low populated and water areas. Water areas were excluded from the analysis as all charging pools are located on a dry land. If values of a predictor that was derived from GIS attribute is missing for only one or a small group of charging pools, either a given attribute or charging pools should be excluded from the analysis as the used methods cannot handle missing data. Hence, imputing only a few missing values may save a lot of useful data for the analysis. We applied the following set of if-then rules to impute missing values in GIS polygons:
\begin{enumerate}[label=\alph*)]
	\item If a polygon is not inhabited, the missing values of attributes informing about population subcategories (e.g. number of persons receiving social assistance or disability benefits) were set to zero. 
	
	\item If the number of income recipients in a polygon equals to zero, then missing values of attributes related to population groups receiving a certain type or level of income were set to zero. 
	
	\item If the number of buildings in a polygon equals to zero, then we set to zero attributes informing about the specific types of constructions (e.g. rental properties, homes owned by authorized institutions or municipalities, the percentage of multi-family housing, the average house value and the average house occupancy). 
	
	\item If the number of households in a polygon is zero, then the missing number of households in various subcategories is set to zero.
	
	\item If the urbanity class in a polygon is missing, then, the lowest urbanity class (non-urban area) is supplied. 
	
	\item If the percentage of addresses in an area with the same zip code is missing, then the lowest value of percentage (less than 50 \% of addresses) is used.
	
	\item If the number of measurement points for electricity or gas is less than five, then missing values of electricity or gas consumption are set to zero.
\end{enumerate}

If applicable, missing values of attributes were estimated from known values of some other attributes, e.g. the missing value of the mean size of household was replaced by the number of inhabitants divided by the number of households. If none of the above-mentioned rules could be applied, the value of a GIS polygon attribute remained missing.

\subsection{Extraction of predictors from GIS data}
\label{secSI:preparation_of_predictors}

To extract predictors from the GIS data for the immediate vicinity of charging pools, a buffer~\cite{Thornton_2011} (i.e. the circular area with a specific radius and centered at the position of a charging pool) was used. To extract the predictors from polygon GIS data, intersections of buffers and polygons were found. The value of an attribute for a buffer area was estimated assuming uniform spatial distribution of the attribute inside the polygon. Based on this assumption, a sum of estimates for buffer intersections was calculated to estimate attribute value for the buffer (e.g. estimate of the population for a buffer was calculated as the sum of populations of all buffer-polygon intersections). For attributes taking relative (e.g. percentage of rental properties in the polygon) or average (e.g. average income per individual receiving an income) values, the buffer value was calculated as a weighted average of attribute values that are associated with polygons intersecting the buffer, while using the area of the intersection as a weight. For instance, average address density in a buffer was estimated as a weighted average of address densities in all buffer-polygons intersections, weighted by areas of intersections. In special cases, to improve the estimates, the values of related attributes were used as weights, when calculating the weighted average over the buffer-polygon intersections. Weighting with values of the related attribute can help to estimate better the resulting value, than weighting with the buffer area, e.g. for average house value a more suitable related attribute is the number of houses. For categorical attributes, e.g. land use, for each category a new predictor was derived, representing the proportion of buffer area filled in by a given land use category.

With polyline data, such as traffic flow data, we identified for each charging pool the closest road segment and used its traffic flow data as predictors. Besides, for each buffer area, we calculated the traffic density by multiplying the length of all road segments by the traffic flow and dividing it by the area of the buffer. Furthermore, the road density was calculated by dividing the length of road segments by the buffer size and used as a predictor~\cite{Liu_2017}. Using one-hot encoding \cite{Lucas2019}, the road segment type (residential, primary, secondary or tertiary) of the closest road segment was added as a predictor.

From the point data, represented by  OpenStreetMap points of interest and Charging pools 2015 datasets, we extracted predictors by calculating the density of points in the buffer and by considering the distance to the closest point. Finally, we derived five predictors from the EVnetNL dataset: the number of connectors, maximum power of a charging pool, latitude, longitude and the rollout of the charging pool (strategic or demand-driven).

To make sure that predictors are not built based on GIS data with a large proportion of missing values, if there was less than 1.5\% of predictor values missing after applying the estimations (see section~\ref{secSI:missing_values_handling}), missing values were replaced by a median value, otherwise, the predictor was discarded. The number of predictors derived from each GIS dataset is reported in Table~\ref{tab:GISdata}.

\subsection{Handling potential data problems}
\label{secSI:handling_data_problems}
When fitting a model to data many problems may arise. Collinearity, a situation when predictors are related to each other, is one of them~\cite{james2013introduction}. Collinearity may hamper the interpretability of a model. One way how collinearity is demonstrated is due to correlations between predictors. To reduce the level of correlations between predictors, a removal of correlated predictors was applied~\cite{kuhn2013applied}. In the first step for all pairs of predictors, the value of the Pearson correlation coefficient was calculated. 

In the second step, we identified groups of predictors with the absolute value of the Pearson correlation coefficient between each pair of features greater than 0.95. For each group, we kept only one out of these correlated predictors as a representative of the group and excluded other predictors. By this step, we have omitted 16 predictors.

To avoid using data that carry none or very little useful information, we analyzed in all predictors the frequency of unique values. We found some cases (in predictors derived from land-use data) when predictors had more than 95\% of values equal to zero. All 10 such predictors were eliminated as they may cause difficulties when splitting data into training and testing data if only zero values are selected into the training dataset.

Consideration of non-linearities in the data may improve the quality of the model. We extended the set of predictors by adding predictors obtained by the application of functions $\log(\cdot)$, $\sqrt{\cdot}$ and $(\cdot)^2$ to all predictors and all pairwise products of predictors (interaction terms) as it is often recommended in the literature~\cite{james2013introduction}. However, these operations did not lead to significant model improvement, so they were excluded from the data pre-processing.

\section{Description of predictors published together with the paper}

The names and descriptions of predictors included in the attached data matrix are described in the Table~\ref{tabSI:predictorDesc}. The response variable, the number of unique RFID cards used on corresponding charging pools, is available in the column denoted as $n_{RFID}$.

\begin{longtable}[h]{p{.2\textwidth} p{.8\textwidth}}
	\hline
	Column & Explanation \\ 
	\hline
	PC2 & Average age within the population core by January 1, 2011. \\ 
	PC3 & Number of recreational properties within the core area. \\ 
	PC4 & Number of individuals aged 15 - 24 in one-person households. \\ 
	PC5 & Number of individuals aged 25 - 44 in one-person households. \\ 
	PC6 & Number of individuals aged 45 - 64 in one-person households. \\ 
	PC7 & Number of individuals aged 65 or more in one-person households. \\ 
	PC8 & Number of individuals aged 0 - 14 in multi-person households without biological or adopted children, or stepchildren. \\ 
	PC9 & Number of individuals aged 15 - 24 in multi-person households without biological or adopted children, or stepchildren. \\ 
	PC10 & Number of individuals aged 25-44 in multi-person households without biological or adopted children, or stepchildren. \\ 
	PC11 & Number of individuals aged 45 - 64 in multi-person households without biological or adopted children, or stepchildren. \\ 
	PC12 & Number of individuals aged 65 or more in multi-person households without biological or adopted children, or stepchildren. \\ 
	PC13 & Number of individuals aged 0 - 14 in multi-person households with biological or adopted children, or stepchildren. \\ 
	PC14 & Number of individuals aged 15 - 24 in multi-person households with biological or adopted children, or stepchildren. \\ 
	PC17 & Number of individuals aged 65 or more in multi-person households with biological or adopted children, or stepchildren. \\ 
	PC18 & Number of individuals living as unmarried couples or unregistered couples without children belonging to a private household of two people. \\ 
	PC22 & Number of individuals in private households of one parent with at least one child living at home. \\ 
	PC24 & Number of individuals inhabiting institutional properties, such as nursing homes, elderly and children's homes, group homes, rehabilitation centres and prisons.  \\ 
	PC25 & Number of individuals whose parents were both born in the Netherlands, regardless of their own homeland. \\ 
	PC26 & Number of individuals with at least one parent born in Europe (excluding Turkey), North America, Oceania, Indonesia or Japan. \\ 
	PC27 & Number of individuals with at least one parent born in Africa, Latin America, Asia (excluding Indonesia and Japan) or Turkey. \\ 
	PC28 & Working population, 25-44 years old [\%]. \\ 
	PC29 & Working population, 45-54 years old [\%]. \\ 
	PC30 & Working population, 55-64 years old [\%]. \\ 
	PC31 & Working population, 65-74 years old [\%]. \\ 
	PC32 & Individuals of the working population working in agriculture, forestry and fishing [\%]. \\ 
	PC33 & Working population in the mining, manufacturing or construction [\%]. \\ 
	PC34 & Percentage of the working population employed in commercial services corresponding to  categories: wholesale and retail, transportation and storage, information and communication, financial services, hire and selling of real estate, lease of movable goods and other business services, veterinary services. \\ 
	PC35 & Percentage of the working population engaged in non-commercial services corresponding to categories: public administration and public services, education, health and welfare, culture, sports and recreation, other services, households as employers, extraterritorial organisations. \\ 
	PC36 & Percentage of the working population engaged in one of the following categories: power, waterworks and waste management, hotel and catering, restaurants, specialist business services, except veterinary services. \\ 
	PC37 & Number of individuals moving into the geographical area minus the number of individuals moving out elsewhere in the Netherlands in the period 1 January 2001-1 January 2011. \\ 
	PC38 & Net number of immigrants in the population core, i.e. the number of immigrants moving into the population core minus the number of individuals who emigrated in the period 1 January 2001-1 January 2011. \\ 
	PC39 & Number of households with two individuals. \\ 
	PC40 & Number of households with three individuals. \\ 
	PC41 & Number of households with four individuals. \\ 
	PC42 & Number of households with five individuals. \\ 
	PC43 & Number of households with six or more individuals. \\ 
	PC44 & Average house value. \\ 
	N\_1 & The residential properties [\%]. The number is stated as a percentage of the total number of dwellings. \\ 	
	N\_3 & Urban class of the neighbourhood, based on the density of properties (five classes). \\ 
	N\_5 & Average income per individual receiving an income. \\ 
	N\_6 & Average income per inhabitant. \\ 
	N\_7 & Percentage of individuals in private households belonging to the nationwide 40\% with the lowest personal income. \\ 
	N\_8 & Percentage of individuals in private households belonging to the nationwide 20\% with the highest personal income. \\ 
	N\_9 & Percentage of private households belonging to the nationwide 40\% of households with the lowest household income. \\ 
	N\_10 & Percentage of private households belonging to the nationwide 20\% of households with the highest household income. \\ 
	N\_11 & Percentage of households with low purchasing power. \\ 
	N\_13 & Percentage of economically inactive individuals aged between 15 and 65 (unemployed, disabled, unemployed students, etc.). \\ 
	N\_14 & Percentage of addresses in the neighbourhood with the most common zip code.  \\ 
	N\_15 & Average number of addresses in a neighbourhood within a one kilometre radius circle (seeks the degree of concentration of  human activities (living, working, going to school, shopping, entertainment, etc.))  \\ 
	N\_16 & Total population. \\ 
	N\_18 & Population aged 15 to 24 years [\%]. \\ 
	N\_19 & Population aged 25 to 44 years [\%]. \\ 
	N\_20 & Population aged 45 to 64 years [\%]. \\ 
	N\_21 & Population aged 65 years or older [\%]. \\ 
	N\_24 & Number of multi-person households without children expressed in whole percentage of the total number of private households. \\ 
	N\_25 & Unmarried residents [\%]. \\ 
	N\_26 & Married residents [\%]. \\ 
	N\_27 & Widowed residents [\%]. \\ 
	N\_29 & Number of multi-family housings as a percentage of the total housing stock. \\ 
	N\_30 & Number of homes built in 2000 or later, expressed as a percentage of the total number of homes. \\ 
	N\_31 & Average household size. \\ 
	N\_32 & Average value of residential real estate property. \\ 
	N\_33 & Number of live births in 2015. \\ 
	N\_34 & Number of live births in 2015 per thousand inhabitants. \\ 
	N\_35 & Number of deaths in 2015. \\ 
	N\_36 & Number of deaths in 2015, per thousand inhabitants. \\ 
	N\_37 & Number of immigrants with a Western origin (Europe (excluding Turkey), North America and Oceania or Indonesia or Japan), expressed as a percentage of the entire population. \\ 
	N\_38 & Number of immigrants with non-Western origin, expressed as a percentage of the entire population. These immigrants belong to the ethnic group one of the countries in the continents of Africa, Latin America and Asia (excluding Indonesia and Japan) or Turkey. \\ 
	N\_39 & Number of immigrants with Morocco origin, expressed as a percentage of the entire population. \\ 
	N\_40 & Number of immigrants with ethnic group of (former) Netherlands Antilles and Aruba, expressed as a percentage of the entire population. \\ 
	N\_41 & Number of immigrants with ethnic group of Suriname, expressed as a percentage of the entire population. \\ 
	N\_42 & Number of immigrants with ethnic group of Turkey, expressed as a percentage of the entire population. \\ 
	N\_43 & Number of immigrants with other non-western origin, expressed as a percentage of the entire population. \\ 
	N\_44 & Number of agriculture, forestry and fishing businesses. \\ 
	N\_45 & Number of industry and energy businesses. \\ 
	N\_46 & Number of wholesale, shops, hotels, restaurants and catering facilities. \\ 
	N\_47 & Number of transportation, information and communication businesses. \\ 
	N\_48 & Number of financial services, real estate businesses. \\ 
	N\_49 & Number of business services. \\ 
	N\_50 & Number of cultural, recreational and other services. \\ 
	N\_51 & Number of business establishments. \\ 
	N\_52 & Number of individuals who receive a disability benefit under the Law on Disability Insurance, the invalidity insurance, the Work and Income according to labour capacity,  the Invalidity Disabled Young Individuals and the work and employment support for Disabled Young individuals Act. \\ 
	N\_53 & Number of individuals receiving benefits under the Unemployment Insurance Act. \\ 
	N\_54 & Number of persons receiving social assistance. \\ 
	N\_55 & Number of motor vehicles for road passenger transport, excluding mopeds and motorcycles, with up to nine seats (including the driver). \\ 
	N\_56 & Number of motor vehicles for road passenger transport, excluding mopeds and motorcycles, with up to nine seats (including the driver). \\ 
	N\_57 & Number of passenger cars per household. \\ 
	N\_58 & Number of vans, lorries, tractors (motor vehicle equipped to tow trailers), special vehicles (commercial vehicles for special purposes such as fire trucks, cleaning cars, tow trucks) and buses.  \\ 
	N\_59 & Number of motorcycles, scooters, motor carriers and motor wheelchairs with a motorcycle registration. \\ 
	N\_60 & Number of cars aged six years and older. \\ 
	N\_61 & Number of cars driving on petrol.  \\ 
	N\_62 & Percentage of homes that are in the possession of an authorised housing institution or a municipal housing company. \\ 
	N\_63 & Percentage of homes that are not owned by a housing association. \\ 
	N\_64 & Percentage  of vacant homes. \\ 
	LC\_1 & Proportion of railway areas. \\ 
	LC\_2 & Proportion of road traffic areas. \\ 
	LC\_4 & Proportion of residential areas. \\ 
	LC\_5 & Terrain for retail and catering industry. \\ 
	LC\_6 & Terrain for public facilities. \\ 
	LC\_7 & Terrain for socio-cultural facilities. \\ 
	LC\_8 & Proportion of business and industrial areas. \\ 
	LC\_11 & Proportion of cemetery areas. \\ 
	LC\_13 & Proportion of building sites. \\ 
	LC\_15 & Proportion of parks and plantations. \\ 
	LC\_16 & Proportion of sports fields. \\ 
	LC\_17 & Proportion of terrain for non-commercial, ornamental and vegetable cultivations. \\ 
	LC\_21 & Proportion of other agricultural land. \\ 
	LC\_22 & Proportion of forests. \\ 
	LC\_24 & Proportion of waters. \\ 
	LC\_25 & Proportion of recreational inland waters. \\ 
	EC\_1 & Mean gas consumption of residential properties [$m^3$]. \\ 
	EC\_2 & Annual gas consumption of residential properties [$m^3$]. \\ 
	EC\_4 & Mean electricity consumption of residential properties [kWh]. \\ 
	EC\_7 & Mean gas consumption of companies [$m^3$]. \\ 
	EC\_8 & Annual gas consumption of companies [$m^3$]. \\ 
	EC\_9 & Number of companies, where the gas consumption was measured. \\ 
	EC\_10 & Mean consumption of electric energy of companies [kWh]. \\ 
	EC\_11 & Annual electric energy consumption of companies [kWh]. \\ 
	L\_1 & Liveability index 2016 - subcategory houses, comprehending factors as type, construction year and ownership of houses, deviation from the national average.  \\ 
	L\_2 & Liveability index 2016 - subcategory residents, comprehending types of families, migration background, unemployment rate and similar population characteristics, deviation from the national average. \\ 
	L\_3 & Liveability index 2016 - subcategory services, consisting mostly of availability of walking distance to various services, deviation from the national average. \\ 
	L\_4 & Liveability index 2016 - subcategory safety, evaluating crime rates of different types, deviation from the national average.  \\ 
	L\_5 & Liveability index 2016 - subcategory environment, embracing vicinity of the neighbourhood, e.g. forests, factories, wind turbines, highways etc., deviation from the national average. \\ 
	L\_6 & Liveability index 2016. \\ 
	LS\_1 & Ambient population (average over 24 hours) distribution. \\ 
	NERST\_nr\_car & Daily number of passenger cars on nearest road segment. \\ 
	NERST\_nr\_mw & Daily number of multivans/busses on nearest road segment. \\ 
	NERST\_nr\_truck & Daily number of trucks on nearest road segment. \\ 
	TRDENS\_nr\_car & Daily traffic density of passenger cars. \\ 
	TRDENS\_nr\_mw & Daily traffic density of multivans/busses.  \\ 
	TRDENS\_nr\_\\	truck & Daily traffic density of trucks. \\ 
	road\_density & Length of roads inside buffer [$m$]. \\ 
	n.accomodation & Number of accomodation related OSM amenities. \\ 
	n.culture & Number of culture related OSM amenities. \\ 
	n.education & Number of work related OSM amenities. \\ 
	n.entertainment & Number of entertainment related OSM amenities. \\ 
	n.family & Number of family related OSM amenities. \\ 
	n.fashion & Number of fashion related OSM amenities. \\ 
	n.food & Number of food related OSM amenities. \\ 
	n.health & Number of health related OSM amenities. \\ 
	n.hobby & Number of hobby related OSM amenities. \\ 
	n.household & Number of household related OSM amenities. \\ 
	n.money & Number of finance related OSM amenities. \\ 
	n.public & Number of public related OSM amenities. \\ 
	n.sport & Number of sport related OSM amenities. \\ 
	n.transportation & Number of transportation related OSM amenities. \\ 
	n.work & Number of work related OSM amenities. \\ 
	min\_dist.\\accomodation & Min. dist. to accomodation related OSM amenity. \\ 
	min\_dist.culture & Min. dist. to culture related OSM amenity. \\ 
	min\_dist.\\education & Min. dist. to work related OSM amenity. \\ 
	min\_dist.\\entertainment & Min. dist. to entertainment related OSM amenity. \\ 
	min\_dist.family & Min. dist. to family related OSM amenity. \\ 
	min\_dist.fashion & Min. dist. to fashion related OSM amenity. \\ 
	min\_dist.food & Min. dist. to food related OSM amenity. \\ 
	min\_dist.health & Min. dist. to health related OSM amenity. \\ 
	min\_dist.hobby & Min. dist. to hobby related OSM amenity. \\ 
	min\_dist.\\household & Min. dist. to household related OSM amenity. \\ 
	min\_dist.money & Min. dist. to finance related OSM amenity. \\ 
	min\_dist.public & Min. dist. to public related OSM amenity. \\ 
	min\_dist.sport & Min. dist. to sport related OSM amenity. \\ 
	min\_dist.\\transportation & Min. dist. to transportation related OSM amenity. \\ 
	min\_dist.work & Min. dist. to work related OSM amenity. \\ 
	n\_of\_nn\_chst & Number of charging pools in the buffer. \\ 
	min\_dist\_chst & Min. dist. to a next charging pool. \\ 
	RoadType\_\\residential &  charging pool associated with residential road. \\ 
	RoadType\_\\secondary & charging pool associated with secondary road. \\ 
	RoadType\_\\tertiary & charging pool associated with tertiary road.  \\ 
	ncon & Number of connectors. \\ 
	max\_power & Maximal charging power [kW]. \\ 
	lat & Latitude of charging pool. \\ 
	lon & Longitude of charging pool. \\ 
	CP\_type & Strategical roll-out of charging pool. \\ 
	\hline
	\caption{Predictors used in the analyses. In the first column is listed the abbreviation used in the provided data file, the second column gives the description of predictors.}
	\label{tabSI:predictorDesc}
	
\end{longtable}

\end{document}